\documentstyle[aps,prl,epsf,multicol]{revtex}

\def\ket#1{| #1\rangle}
\def\bra#1{\langle #1 |}
\newtheorem{hypothesis}{Hypothesis}

\title{
\rightline{ \small DAMTP-2001-62\normalsize } \rightline{}
\centerline{Nonsequential positive-operator-valued measurements on}\centerline{entangled mixed states do not always violate a Bell inequality}}
\author{Jonathan Barrett}
\address{Centre for Mathematical Sciences,
Wilberforce Road, Cambridge CB3 0WA, UK}

\begin{document}
\maketitle

\begin{abstract}

We present a local-hidden-variable model for positive-operator-valued
measurements (an LHVPOV model) on a class of entangled generalized Werner
states. We also show that, in general, if the state $\rho'$ can be
obtained from $\rho$ with certainty by local quantum operations without
classical communication, then an LHVPOV model for the state $\rho$
implies the existence of such a model for $\rho'$.   
\vskip10pt
PACS number(s): 03.67.-a, 03.65.Ta, 03.65.Ud
\end{abstract}
\vskip10pt

\begin{multicols}{2}

\section{Introduction}\label{introduction}

It is well known that some quantum states of joint systems are
``nonlocal,'' meaning that outcomes of measurements performed separately
on each subsystem at spacelike separation cannot be reproduced by a
local-hidden-variable (LHV) model (see \cite{peresbook} and references
contained therein). Such nonlocality can
be revealed by a violation of an inequality which any LHV model must satisfy. We call any such inequality a ``Bell-type
inequality.'' More specifically, consider a bipartite state $\rho$
which acts on ${\cal H}_A\otimes {\cal H}_B$ (in this paper, we only
consider bipartite states). The two subsystems are spatially
separated, one being in the possession of an observer Alice and the
other in possession of an observer Bob. If Alice performs a
measurement $A$ with an outcome $A_i$ and, at spacelike separation, Bob
performs a measurement $B$ with an outcome $B_j$, then an LHV model supposes that the joint probability of getting $A_i$
and $B_j$ is given by
\begin{equation}\label{lhvmodel}
\mathrm{Pr}\mathnormal(A_i,B_j|A,B,\rho )=\int\! d\lambda \, \omega^{\rho }(\lambda ) \,
\mathrm{Pr}\mathnormal(A_i|A,\lambda ) \, \mathrm{Pr}\mathnormal(B_j|B,\lambda ),
\end{equation}
where $\omega^{\rho }(\lambda )$ is some distribution over a space,
$\Lambda$, of hidden states $\lambda$. If a Bell-type inequality is
violated, then no such model exists.

It is also well known that any entangled pure state will
violate some Bell-type inequality and is therefore nonlocal \cite{gp,pr}. This
nonlocality can always be revealed by an appropriate choice of
projective measurements to be performed on each subsystem. In
light of this, one might conjecture that the same holds true for mixed
states, namely that with an appropriate choice of projective measurements,
some Bell-type inequality will be violated. The conjecture, however,
is false. That it is false was shown by Werner, who wrote down an
explicit LHV model for projective measurements
performed by Alice and Bob
on a class of mixed entangled bipartite states, now known as ``Werner
states'' \cite{werner,mermin} (in fact, he did this before the results
of \cite{gp,pr} were known). The situation became more complicated
when Popescu showed that certain of the Werner states (specifically those in
${\cal H}_d\otimes {\cal H}_d$, where $d\geq 5$) have a ``hidden
nonlocality'' \cite{hiddennonloc}. He showed that if Alice and Bob perform a sequence of
measurements consisting of a fixed initial projection onto a
two-dimensional subspace followed by a projective measurement
(corresponding to a test of the CHSH inequality \cite{chsh} ``within'' that
subspace) then no LHV model will reproduce the results
correctly. (More exactly, no ``causal'' LHV model can reproduce the results
correctly, where ``causal'' means that the outcome of Alice's first
measurement cannot depend on
her choice of which measurement to perform second.) Teufel \emph{et al.}
address the question of classifying different types of nonlocality in
some detail \cite{teufel} (see also \cite{horosnonloc}). In particular, they demonstrate how some states might only
display what they call ``deeply hidden nonlocality.'' They also give
conditions which causal local models have to satisfy that are more
involved than that of Eq. (\ref{lhvmodel}). Other
investigations include
\cite{gisin} and \cite{pereshiddennonloc}.

It is clear from the above that, regarding the relationship
between entanglement and nonlocality, the situation is rather more
complicated than one might suppose simply from a study of pure
states. In considering nonlocality, we have to consider separately the
cases in which Alice and Bob can perform positive-operator-valued (POV) measurements on their
subsystems and in which they are restricted to projective
measurements. We must also consider whether they are allowed sequences
of measurements or single measurements only and whether these
measurements can be collective, i.e., joint measurements performed on
several particle pairs at once, or are resticted to measurements
performed separately on each particle pair. In this work, we consider
the case in which POV measurements are allowed but Alice and Bob
cannot perform sequences of measurements or collective measurements.

A rather natural sounding hypothesis then emerges. It is hinted at by
Popescu \cite{hiddennonloc} and raised explicitly by Teufel \emph{et al} \cite{teufel}:
\begin{hypothesis}
Any entangled quantum state will violate some
Bell-type inequality if Alice and Bob can perform single (that is,
nonsequential) POV measurements on individual copies of the state.
\end{hypothesis}
We show that this hypothesis is false via the
construction of an explicit LHV model for POV measurements (an ``LHVPOV model'') on a class
of generalized Werner states. The model as presented simulates the state
\begin{equation}\label{rho}
\rho = \alpha \, \frac{2 P^{anti}}{d(d-1)} + (1-\alpha )\frac{I}{d^2},
\end{equation}
where 
\begin{equation}\label{alpha}
\alpha = \frac{1}{d+1}(d-1)^{d-1}d^{-d}(3d-1).
\end{equation}
Here, $I$ is the
identity in ${\cal H}_d\otimes {\cal H}_d$ and $P^{anti}$ projects
onto the antisymmetric subspace. The state $\rho$ is
entangled if and only if $\alpha > 1/(1+d)$
\cite{werner}. With $\alpha$ defined by Eq. (\ref{alpha}), $\rho$ is
entangled for any $d\geq 2$. The states originally introduced by
Werner were of the form of $\rho$ but with $\alpha$ set to $(d-1)/d$.

We present the model in Sec. \ref{model}. In
Sec. \ref{extension} we show that this model implies the existence of
an LHVPOV model for a wide class of other entangled mixed
states. Sec. \ref{conclusion} concludes.

\section{The Model}\label{model}

\subsection{Description}\label{description}

In constructing the model, we take some inspiration from Werner's original model for projective
measurements \cite{werner} (it was also inspired by the models of
\cite{cerf} and \cite{massar}). The hidden state is a vector in
$d$-dimensional complex Hilbert space, which we denote by $\ket{\lambda
}$. The distribution of $\ket{\lambda }$ states, $\omega(\lambda )$, is invariant under
$U(d)$ rotations. Note that $\ket{\lambda }$ is a hidden state, not a
quantum state; we write it as a ket merely for convenience. A hidden
state $\ket{\lambda }$ defines probabilities for Alice's and Bob's
measurement outcomes. First, we define rules which work in the case
that all POVM elements are proportional to projectors. At the end
of this section, we will show that this model implies fairly trivially
the existence of a model for all POV measurements. We suppose, then, that
Alice performs a measurement $A$, corresponding to a decomposition of
the identity $\sum_i A_i = I$, where $A_i = x_i P_i$, $0\leq x_i\leq
1$, and $P_i$ is a projection operator. Similarly, Bob performs a
measurement $B$, where $B_j = y_j Q_j$.
\vskip10pt
{\bf Alice.} Restrict attention to those $A_i$ such that $\bra{\lambda }P_i \ket{\lambda } > 1/d$.
Either exactly one of these $A_i$ will be ``accepted'' or ``rejection'' will
occur. The probability of $A_i$ being accepted is given by
$\bra{\lambda}A_i\ket{\lambda}$. If $A_i$ is accepted, the
corresponding measurement outcome is obtained. If no $A_i$ is accepted, then rejection has occurred. In this
case, we widen our attention again to the complete set of $A_i$ and
outcome $i$ is obtained with probability $x_i/d$.

It follows that
\begin{eqnarray}\label{alice}
\lefteqn{\mathrm{Pr}(A_i|A,\lambda ) = } \hspace{20pt} \nonumber \\
& & \bra{\lambda }A_i\ket{\lambda} \ \Theta \!\left(\bra{\lambda
}P_i\ket{\lambda }-1/d \right) \nonumber \\
& & {} + \left(1-\sum_k\bra{\lambda }A_k\ket{\lambda
}\ \Theta \!\left(\bra{\lambda }P_k\ket{\lambda }-1/d \right)\right)\frac{x_i}{d},
\end{eqnarray}
where $\Theta$ is the Heaviside step function.
\vskip10pt
{\bf Bob.} Define
\begin{equation}\label{bob}
{\rm Pr}(B_j|B,\lambda ) = \frac{1}{d-1}y_j\left(1-\bra{\lambda }Q_j\ket{\lambda }\right).
\end{equation}
\vskip10pt
Substituting Eqs. (\ref{alice}) and (\ref{bob}) into Eq. (\ref{lhvmodel}), we get
\begin{eqnarray}\label{correlation}
\lefteqn{{\rm Pr}(A_i,B_j|A,B,\rho ) = } \hspace{20pt} \nonumber \\
&& \hspace{-10pt} \int \! d\lambda \ \omega(\lambda ) \, \bigg[\bra{\lambda }A_i\ket{\lambda} \ \Theta \!\big(\bra{\lambda
}P_i\ket{\lambda }-1/d\big) \nonumber \\
& & \hspace{20pt} + \bigg(1-\sum_k\bra{\lambda }A_k\ket{\lambda
}\ \Theta \!\big(\bra{\lambda }P_k\ket{\lambda
}-1/d\big)\bigg)\frac{x_i}{d} \bigg] \nonumber \\
&& \hspace{20pt} \times \, \frac{1}{d-1}y_j\left(1-\bra{\lambda }Q_j\ket{\lambda }\right).
\end{eqnarray}
We aim to show that this is equal to the quantum prediction: ${\rm Tr}
\; (\rho \, A_i\otimes B_j)$. 

\subsection{Proof that the model works}\label{proof}

We define
\begin{equation}\label{defineJ}
J_{ij}\equiv x_iy_j\!\int\!\!{\rm d}\lambda \,\omega(\lambda ) \,\Theta\!\left(\bra{\lambda
}P_i\ket{\lambda }-1/d\right) \,\bra{\lambda }P_i\ket{\lambda } \,
\bra{\lambda }Q_j\ket{\lambda }.
\end{equation}
We can write Eq. (\ref{correlation}) as
\begin{eqnarray}
\lefteqn{{\rm Pr}(A_i,B_j|A,B,\rho )} \hspace{3pt} \nonumber \\
&=& \frac{1}{d-1}\left(-J_{ij} - \frac{1}{d}x_iy_j\!\int\!\!{\rm d}\lambda \,\omega(\lambda)\,\bra{\lambda }Q_j\ket{\lambda
} \right) \nonumber \\
&& + \frac{1}{d-1}\left(\frac{x_i}{d} \sum_k J_{kj} \right) \nonumber \\
&&+\frac{y_j}{d-1}\sum_l\left(J_{il} + \frac{1}{d}x_iy_l\!\int\!\!{\rm d}\lambda \,\omega(\lambda)\,\bra{\lambda }Q_l\ket{\lambda
}\right)\nonumber \\
&& - \frac{y_j}{d-1}\sum_l\left(\frac{x_i}{d} \sum_k J_{kl} \right) \nonumber \\
&=& \frac{1}{d^2}x_iy_j + \frac{1}{d-1}\left(-J_{ij} +\frac{x_i}{d}
\sum_k J_{kj}\right) \nonumber \\
&& + \frac{1}{d-1}\left(y_j\sum_lJ_{il}-\frac1d x_iy_j\sum_{k,l}J_{kl}\right). \label{correlation2}
\end{eqnarray}
We have used the fact that $\sum_jy_jQ_j=I$.

It remains to calculate $J_{ij}$.
Following Mermin \cite{mermin}, we write $\ket{\lambda } =
\sum_{\nu=1}^d z_{\nu } \ket{\nu }$, where the $\ket{\nu }$ are an
orthonornal basis and $z_{\nu} = r_{\nu} e^{i {\theta_{\nu}}}$. Our
strategy will be to choose coordinates such that $P_i =
\ket{1}\bra{1}$ and, again following Mermin, to substitute $u_{\nu} = r_{\nu}^2$.

We get
\begin{eqnarray}
J_{ij} &=& x_i y_j \!\int \!\! {\rm d}\lambda \ \omega(\lambda )
\, \Theta\!\left(\bra{\lambda }P_i\ket{\lambda } - 1/d \right) \,
\bra{\lambda }P_i\ket{\lambda } \, \bra{\lambda }Q_j\ket{\lambda }
\nonumber \\
       &=& \frac1N \ x_i \ y_j \ \sum_{\nu =1}^{d} \
|\langle q_j | \nu \rangle |^2 \nonumber \\
&& \hspace{-10pt} \times \int_{\frac1d}^{1}\!{\rm d}u_1\int_0^1\!{\rm
d}u_2\ldots \int_0^1\!{\rm d}u_d \,
\delta(u_1+\cdots +u_d-1) \,u_1 u_{\nu } \nonumber \\
       &=& x_i \ y_j \ \sum_{\nu =1}^{d} \ |\langle q_j | \nu
\rangle |^2 \ J_{\nu },
\end{eqnarray}
where
\begin{equation}
N = \int_0^1\!{\rm d}u_1\ldots {\rm d}u_d \ \delta(u_1+\cdots +u_d-1),
\end{equation}
\begin{equation}
Q_j = \ket{q_j}\bra{q_j},
\end{equation}
and
\begin{eqnarray}
\lefteqn{J_{\nu } = \frac1N \int_{\frac1d}^{1}\!\!{\rm d}u_1\int_0^1\!\!{\rm
d}u_2\ldots} \hspace{35pt} \nonumber \\
&& \ldots \int_0^1\!\!{\rm d}u_d \,
\delta(u_1+\cdots +u_d-1) \, u_1 \, u_{\nu }\label{jnu}.
\end{eqnarray}
We can use the fact that for $\nu =2,\ldots ,d$,
\begin{equation}
J_{\nu } = \frac{1}{d-1}(J_2+\cdots +J_d) = \frac{J_0 - J_1}{d-1},
\end{equation}
where $J_0$ is defined by Eq. (\ref{jnu}), setting $u_0 = 1$, and
\begin{equation}
\sum_{\nu =2}^{d} \ |\langle q_j | \nu \rangle |^2 = 1 - |\langle q_j | 1
\rangle |^2,
\end{equation}
giving
\begin{equation}
J_{ij} = x_i \ y_j \ \left( J_1 |\langle q_j | 1 \rangle |^2
+ \frac{J_0 - J_1}{d-1} \left( 1 - |\langle q_j | 1 \rangle |^2 \right)
\right).
\end{equation}

Finally, we have chosen $\ket{1}$ so that $\ket{1}\bra{1} = P_i$, so
instead of $\ket{1}$ we now write $\ket{p_i}$:
\begin{eqnarray}
J_{ij} &=& x_i \ y_j \ \left( J_1 |\langle p_i | q_j \rangle |^2
+ \frac{J_0 - J_1}{d-1} \left( 1 - |\langle p_i | q_j \rangle |^2 \right)
\right) \nonumber \\
  &=& x_i\ y_j\ \frac{J_0 - J_1}{d-1} + \alpha \ x_i\ y_j\ \frac{|\langle p_i | q_j \rangle
  |^2}{d}, \label{firstterm}
\end{eqnarray}
where
\begin{equation}
\alpha = \frac{d^2J_1-dJ_0}{d-1}.
\end{equation}
In calling this quantity $\alpha$, we are anticipating the fact that
it will turn out to be equal to the $\alpha$ of Eqs. (\ref{rho})
and (\ref{alpha}). 

Plugging Eq. (\ref{firstterm}) into Eq. (\ref{correlation2}), the expression
for the correlation predicted by the model, we get, after some algebra
and using the facts that $\sum_i x_iP_i=I$ and $\sum_ix_i=d$,
\begin{eqnarray}
\lefteqn{{\rm Pr}(A_i,B_j|A,B,\rho )} \hspace{2pt} \nonumber \\
&=&
\left(\frac{d-1+\alpha}{d^2(d-1)}\right)x_iy_j-\frac{\alpha}{d(d-1)}|\langle
p_i|q_j\rangle|^2x_iy_j. \label{correlation3}
\end{eqnarray}

It is easy to show that this is in fact equal to the quantum
prediction, ${\rm Tr}(\rho \, A_i\otimes B_j)$,  for a generalized
Werner state, as defined in Eq. (\ref{rho}) (see,
for example, \cite{werner,mermin}).
The task now is to find $\alpha $. To this end, we need to
evaluate $J_0$ and $J_1$. Here we simply state the results:
\begin{eqnarray}
J_0 &=& \frac1d\left(1-\frac1d\right)^{d-1} +
\frac1d\left(1-\frac1d\right)^d \label{j0result} \\
J_1 &=&
\Bigg[\left(\frac1d\right)^2+\frac{2}{d^2}\left(1-\frac1d\right)
+\frac{2}{d(d+1)}\left(1-\frac1d\right)^2\Bigg] \nonumber \\
&& \hspace{30pt} {}\times \left(1-\frac1d\right)^{d-1}.
\label{j1result}
\end{eqnarray}
This gives, as promised,
\begin{equation}
\alpha = \frac{1}{d+1}(d-1)^{d-1}d^{-d}(3d-1).
\end{equation}

There is one thing left to do, which is to show that an LHV model which
works when the positive operators are proportional to projectors
implies the existence of a model which works for all POV measurements.
This follows from the spectral decomposition theorem. Any POVM
element, $A_i$, satisfies $A_i=A_i^{\dagger}$ and $0\leq A_i \leq
1$. It follows that we can write $A_i=\sum_j c_{ij}P_{ij}$, where the
$c_{ij}$ are real constants such that $0\leq c_{ij} \leq 1$ and the
$P_{ij}$ are one-dimensional projection operators satisfying
$P_{ij}P_{ij'}=\delta_{jj'}P_{ij}$. If each $A_i$ is written in this
form, then we can regard our observer as performing a more ``fine-grained'' POV measurement than the one they actually perform, with
elements $c_{ij}P_{ij}$, and our model will
make appropriate predictions. If the outcome $P_{ij}$ is predicted by
the model, then we can say that outcome $A_i$ is actually
obtained. The only remaining wrinkle arises when we consider that we
may sometimes have $c_{ij}=c_{ij'}$, where $j\neq j'$. In this case, the
spectral decomposition for the operator $A_i$ is not
unique. We get around this problem by including in the specification of the LHV model a specification of
a map from each such $A_i$ to one of its valid spectral
decompositions. The choice of map is arbitrary but must remain fixed
for each run of the Bell-type experiment being simulated. (A
similar manoeuvre is required in the case of Werner's LHV model for
projective measurements on Werner states if we want to be able to
predict outcomes for degenerate projective measurements \cite{werner,mermin}.)

\section{Extending the model}\label{extension}

It is interesting to investigate which other entangled states might
admit an LHVPOV model. In fact, one can show that,
quite generally, an LHVPOV model for the state
$\rho_1$ implies the existence of an LHVPOV model for
the state $\rho_2$ if
\begin{equation}
\rho_2=\sum_{ij} M_i\otimes N_j \rho_1 M_i^{\dagger}\otimes
N_j^{\dagger},
\end{equation}
where $\sum_i M_i^{\dagger}M_i=I$, $\sum_j N_j^{\dagger}N_j=I$, and $I$
is the identity. Equivalently, an LHVPOV model for $\rho_1$ implies
the existence of an LHVPOV model for $\rho_2$ if $\rho_2$ can be
obtained from $\rho_1$ with certainty by local operations (without
classical communication). To show this, call the LHVPOV model for $\rho_1$ ``model
1.'' We aim to define an LHVPOV model (``model 2'') for the state
$\rho_2$. We denote probabilities assigned by model 1 by ${\rm
Pr}^1(\ldots )$ and those assigned by model 2 by ${\rm Pr}^2(\ldots
)$. Models 1 and 2 will involve the same space of hidden states and
the same distribution, $\omega(\lambda )$, over hidden states. We define ${\rm
Pr}^2(A_i|A,\lambda ) \equiv {\rm Pr}^1(A_i'|A',\lambda
)$, where $A_i' \equiv \sum_k M_k^{\dagger}A_iM_k$ and ${\rm
Pr}^2(B_j|B,\lambda ) \equiv {\rm Pr}^1(B_j'|B',\lambda )$, where $B_j' \equiv
\sum_l N_l^{\dagger}B_jN_l$. The $A_i'$ form a decomposition of the
identity and we denote the corresponding measurement by $A'$
(similarly $B_j'$ and $B'$). This ensures that model 2 will make the
correct predictions for $\rho_2$ because
\begin{eqnarray}
\lefteqn{\int\!\!{\rm d}\lambda \, \omega(\lambda )\, {\rm
Pr}^2(A_i|A,\lambda )\, {\rm Pr}^2(B_j|B,\lambda )} \nonumber \\
&=& {\rm Tr}\left(A_i'\otimes B_j' \right)\rho_1 \nonumber \\
&=& \sum_{kl}{\rm Tr}\left(M_k^{\dagger}A_iM_k\otimes
N_l^{\dagger}B_jN_l \right)\rho_1 \nonumber \\
&=& \sum_{kl}{\rm Tr}\left(A_i \otimes B_j \right)\left(M_k\otimes
N_l\rho_1M_k^{\dagger}\otimes N_l^{\dagger}\right) \nonumber \\
&=& {\rm Tr} (A_i\otimes B_j)\rho_2.
\end{eqnarray}

\section{Conclusion}\label{conclusion}

Nonlocality is one of the distinctly nonclassical features of quantum
mechanics. In some situations we might view the nonlocality of a
quantum state as a resource in much the same way that entanglement is
now viewed as a resource. The nonlocality of quantum states thus
deserves an investigation paralleling the work done on the
quantification and manipulation of entanglement. In addition, we might
investigate the relationships between entanglement and nonlocality.
 
To this end, we have presented a model which simulates arbitrary single POV measurements on
single copies of a class of (entangled) generalized Werner states. The
hypothesis that any entangled state has
nonlocality which can be revealed by single POV measurements on
individual copies is thus false. A natural hypothesis which remains
unknown is:
\begin{hypothesis}
Any entangled quantum state can be shown to be nonlocal if arbitrary
sequences of POV measurements are allowed on individual copies of the
state.
\end{hypothesis}
It might be interesting to try to prove this hypothesis false by
extending the model above to sequences of measurements.

\vskip5pt

\leftline{\bf Acknowledgments}

I am grateful to Trinity College, Cambridge for support, CERN for
hospitality, and to the European grant EQUIP for partial support.
I would like to thank Adrian Kent for useful discussions.

\end{multicols}
\end{document}